\documentclass[prl,showpacs,twocolumn,typeset,superscriptaddress]{revtex4}
\usepackage{amsfonts}
\usepackage{amsmath}
\usepackage{amssymb}
\usepackage{graphicx}
\usepackage{epsfig}
\usepackage{citesort}
\usepackage{dcolumn}
\usepackage{bm}

\begin{document}

\title{Optimal quantum state reconstruction for cold trapped ions}
\author{A.B. Klimov}
 \affiliation{Departamento de Física,
Universidad de Guadalajara, Revolución 1500, 44410, Guadalajara,
Jal., México}
\author{C. Mu\~noz}
\affiliation{Departamento de Física, Universidad de Guadalajara,
Revolución 1500, 44410, Guadalajara, Jal., México}
\author{A. Fern\'andez}
\affiliation{Center for Quantum Optics and Quantum Information,
Departamento de Física, Universidad de Concepción, Casilla 160-C,
Concepción, Chile}
\author{C. Saavedra}
\affiliation{Center for Quantum Optics and Quantum Information,
Departamento de Física, Universidad de Concepción, Casilla 160-C,
Concepción, Chile}
\date{Dated: March 12, 2008}

\begin{abstract}
We study the physical implementation of an optimal tomographic
reconstruction scheme for the case of determining the state of a
multi-qubit system, where trapped ions are used for defining
qubits. The protocol is based on the use of mutually unbiased
measurements and on the physical information described in H.
H\"{a}ffner \emph{et. al} [Nature \textbf{438}, 643-646 (2005)].
We introduce the concept of physical complexity for different
types of unbiased measurements and analyze their generation in
terms of one and two qubit gates for trapped ions.
\end{abstract}

\pacs{03.67.Lx, 03.65.Wj, 03.65.-w}
\maketitle

A main task in any experimental physical setup for implementing
quantum computation is the ability to determine the output state
of any given quantum algorithm \cite{Bennett2000}. The standard
procedure applied for quantum state reconstruction of a density
operator lying in a $2^{N}$ dimensional quantum system, in the
case of $N$ qubits, consists in projecting the density operator
onto $3^{N}$, completely factorized, bases in the corresponding
Hilbert space \cite{Fano1957}. All these measurements are obtained
by applying rotations on single qubits (which are referred to as
local operations) followed by projective measurements onto the
logical basis. This was recently achieved for the case of eight
qubits, with trapped ions \cite {Blatt2005}. The experiment was
done by following the quantum computer architecture based on ions
in a linear trap proposed by Cirac and Zoller \cite{Cirac1995}.
Besides, the experimental implementations of several quantum
protocols have also been reported by using trapped ions
\cite{Blatt2004}. In all these cases the quality of the protocols
is tested using standard tomography for quantum state
determination. This scheme has also been used in the cases of
considering optical setups \cite{Mitchell2003} and NMR
\cite{Bouland2003}.

As it was mentioned above, in the standard measurement scheme only
local operations are required to generate all the necessary
projections. In each basis (setup) $2^{N}-1$ independent
measurements can be performed, so that not all the experimental
outcomes obtained in different bases are linearly independent,
that is, there are redundant measurements. In the case of a
$N$-qubit system the anti-diagonal elements have the larger
errors. Actually, accumulated errors are not uniform; these errors
depend on the number of single logic gates used for determining
given elements, so that larger errors appear when single logic
gates act on all the particles. Assuming that there is an error
$\varepsilon $ in the measurement of ion populations, then the
accumulated error for anti-diagonal elements is of the order of
$\varepsilon \sqrt{2^{N-1}+2^{N-2}(2^{N}-1)}$. These errors may
lead to a density operator which does not satisfy the positiveness
condition and so the information from the experimental data must
be optimized. For this purpose the maximum likelihood estimation
(MLE) method \cite{Hradil1997} has been used for the improvement
of the density operators in experiments with light qubits
\cite{James2001} as well as in experiments with matter qubits
\cite{Blatt2004}.

It is well known that the optimal quantum state determination is related to the concept
of measurements on Mutually Unbiased Bases (MUB) \cite{Wootters1989}, which we will
simply refer to as MUB-tomography. Such bases possess the property of being maximally
incompatible. This means that a state producing precise measurement results in one set
produces maximally random results in all the others. The set of mutually unbiased
projectors given by
\begin{equation}
P_{n}^{(\alpha )}=|\psi _{n}^{(\alpha )}\rangle \langle \psi _{n}^{(\alpha
)}|,\;n=1,...,2^{N},\;\alpha =1,..,2^{N}+1,  \label{projector}
\end{equation}%
where
\begin{equation}
Tr\left( P_{n}^{(\alpha )}P_{n^{\prime }}^{(\alpha ^{\prime })}\right)
=\delta _{\alpha \alpha ^{\prime }}\delta _{nn^{\prime }}+\frac{1}{2^{n}}%
(1-\delta _{\alpha \alpha ^{\prime }})
\end{equation}
and $\sum_{n=1}^{2^{N}}P_{n}^{(\alpha)}=I$ ($I$ denotes the
identity), defines a complete set of projection measurements, in
the sense that the measured probabilities
$p_{\alpha_n}=Tr(P_{n}^{(\alpha )}\rho )$ completely determine the
density operator of the system:
\begin{equation}
\rho =\sum_{\alpha =1}^{2^{N}+1}\sum_{n=1}^{2^{N}}p_{\alpha n}P_{n}^{(\alpha )}-I.
\label{rho_ot}
\end{equation}%
The number of MUBs for $N$-qubits is $(2^{N}+1)$, which is
essentially less than $3^{N}$. The use of MUBs can represent a
considerable reduction in the time needed for performing the full
state determination. For instance, recently the reconstruction of
a quantum state codified in the inner states of eight ions in a
linear trap was reported \cite{Blatt2005}. In this experiment the
reconstruction process takes more than $10$ hours, because of the
measurement in $6561$ different bases and a hundred of times for
each one, so the number of measurements is about 656,000, which
quickly grows for an increasing number of qubits. In the case of
using MUB-tomography, the number of measurement bases is only 257
for determining all the elements of the density operator
associated with this state. This could reduce the experimental
time, roughly speaking, to 25 minutes only, where we have
considered a linear interpolation. Recently, an alternative
reconstruction scheme has proposed, which is based on using
pyramidal states for single qubits \cite{Englert2004,Ling2006}.
The implementation of reconstruction using pyramidal states
requires measurement of $N$-particle correlations \cite{Ling2006}.

In the case of MUB-tomography each coefficient $p_{\alpha n}$ in
Eq. (\ref{rho_ot}) has an error associated with the measurement of
only one projector $P_{n}^{(\alpha )}$ (\ref{projector}). Hence,
the error in each coefficient is essentially determined by the
ability of projecting the system onto $P_{n}^{(\alpha )}$. In
practice, such measurements are implemented by projecting the
system onto the logical basis after performing a set of unitary
transformations. Such transformations, due to their nonlocal
features, can be decomposed into a sequence of single and nonlocal
gates, so that the error in this reconstruction is mainly
associated with the quality of these logic gates. Of course, the
reconstructed density operator can be subjected to the MLE method.

The main shortcut of the MUB-tomography scheme is the experimental
implementation of MUB projectors, which is related to the fact
that any set of MUBs contains non-factorizable bases. The
measurements on such non-factorizable bases require application of
non-local gate operations, which currently can not be performed
with fidelity 1. Hence, the natural question for optimizing the
experimental implementation can be casted as: Which is the set of
MUBs that requires the minimum number of nonlocal operations?

To approach this problem, we consider a complete set of MUBs where
the basis factorization is denoted by the following set of natural
numbers: $(k_{1},k_{2},...,k_{\phi (N)})$, where $\phi (n)$ is the
number of possible decompositions of $2^{N}+1$ as a sum of
positive numbers, such that $\sum_{j}k_{j}=2^{N}+1$. Here, the
notation $(k_{1},k_{2},...)$ means that there are $k_{1}$
completely factorized bases; $k_{2}$ bases with two particle
entanglement and all the other particles are factorized, etc. Of
course, for a given number of qubits, only a certain factorization
structure is admitted. For instance, in the two-qubit case the
only allowed structure is (3,2), which means that there are 3
completely factorized bases and 2 non-factorized ones. In the case
of 3 qubits, the bases can be described by the following notation
$(n_{f},n_{b},n_{nf})$, with $n_{f}+n_{b}+n_{nf}=2^3+1$. Here,
$n_{f}$ denotes the number of completely factorized bases; $n_{b}$
the number of bases with bipartite entanglement, i.e., each state
of such bases is factorized as $|\psi \rangle _{ij}|\varphi
\rangle _{k}$,$i\neq j\neq k$, where $i,j,k=1,2,3$; and $n_{nf}$
represents the number of non factorized. For $3$ qubits, there are
4 different sets of MUBs \cite{Lawrence2002,Klimov2005a}, which
are denoted as $(3,0,6)$, $(1,6,2)$, $(2,3,4)$ and $(0,9,0)$.
Because any $N$-qubit entanglement operation can be decomposed
into a sequence of single qubit gates and controlled-NOT gates
(CNOT gates) \cite{Monroe1995}, each basis (in the given set of
MUBs) can be characterized by the minimum number CNOT gates.

In this work we shall concentrate on the experimental
implementation of MUB-tomography in the case of trapped ions as
reported in Ref. \cite{Blatt2004}. In the case of $^{40}Ca^{+}$
trapped ions, where the qubit is codified in the ground state
$\left\vert 0\right\rangle \equiv S_{^{1/2}}(m=-1/2)$ and the
metastable $\left\vert 1\right\rangle \equiv D_{5/2}(m=-1/2)$
state, single logic gates are implemented with a fidelity, $\Phi
_{\mathrm{SG}}$, higher than $99\%$. However, non-local operations
are less accurate; here we will assume the reported value for the
CNOT gate fidelity, $\Phi _{\mathrm{CNOT}}$, which reaches a value
up to $92(6)\%$ \cite{Blatt2004}. Then we can characterize the
\emph{physical complexity} of each set of MUBs as a function of
the number of non-local gates needed for implementing the
projection measurements. Then, a complexity of a given MUB is
characterized by the number $C_{j}$, which can be defined as:
\begin{equation}
C_{j}=\ln \frac{1}{\Phi _{\mathrm{CNOT}}^{n_{j}}}\propto n_{j},
\end{equation}%
where $n_{j}$ is the number of CNOT gates needed for generation of
such a basis. The total complexity of a given set of MUBs is then
\begin{equation}
C=\sum_{j}C_{j}.
\end{equation}%
Thus the total complexity is proportional to the total number of
CNOT gates for preparing all the required projection measurements.
As a simplest example let us consider the case of 3 qubits, where
there are only 4 different sets of MUBs. The bases labeled by
$(n_{f},n_{b},n_{nf})$ have complexities $C\propto
0*n_{f}+1*n_{b}+2*n_{nf}$. This means that the most adequate set
for MUB-tomography is $(0,9,0)$. We have assumed, in this
derivation, that the fidelity of the factorized bases is one, in
the sense that a sequence of single ion gates are required for
their generation. We also assume that CNOT gates between
neighboring ions have the same fidelity as between ions that are
further apart, because CNOT gates are implemented by using the
center of mass motion as a data bus. The decomposition of the
corresponding bases can be generated by starting from the standard
computational basis is that given in Table \ref{opt3}.

\begin{table}[tbp]
\caption{Decomposition of the MUBs in the case of 3-qubits for $(0,9,0)$ structure, where
$R_{k}^{(j)}=R_{k}^{(j)}\left( \protect\pi /2\right)$, with $k=x,y,z$ are the single ions
operations \protect\cite{transfor} and $\chi_{\textrm{CNOT}}^{(ij)}$ is the
controlled-NOT gate with the $\protect\iota $-th and $j$-th ions as source and target,
respectively.} \label{opt3}
\begin{tabular}{|c|c|}
\hline\hline
Basis & Gate operations \\ \hline\hline
1 & $R_{y}^{\left( 1\right) } R_{x}^{\left( 3\right) }
\chi_{\textrm{CNOT}}^{(23)}R_{y}^{\left( 2\right) } $ \\
2 & $R_{x}^{\left( 1\right) } R_{y}^{\left( 3\right) }
\chi_{\textrm{CNOT}}^{(23)}R_{y}^{\left( 2\right) } $ \\
3 & $R_{z}^{\left( 3\right) } \chi_{\textrm{CNOT}}^{(23)}R_{y}^{\left(2\right) }$ \\
4 & $R_{y}^{\left( 3\right) } \chi_{\textrm{CNOT}}^{(13)}R_{y}^{\left(1\right) } $ \\
5 & $R_{x}^{\left( 2\right) } R_{x}^{\left( 3\right) }
\chi_{\textrm{CNOT}}^{(13)}R_{y}^{\left( 1\right) } $ \\
\hline
\end{tabular}
\begin{tabular}{|c|c|}
\hline\hline
Basis & Gate operations \\ \hline\hline
6 & $R_{y}^{\left( 2\right) } R_{z}^{\left( 3\right) }
\chi_{\textrm{CNOT}}^{(13)}R_{y}^{\left( 1\right) } $ \\
7 & $R_{z}^{\left( 2\right) } R_{x}^{\left( 3\right) }
\chi_{\textrm{CNOT}}^{(12)}R_{y}^{\left( 1\right) } $ \\
8 & $R_{y}^{\left( 2\right) } R_{y}^{\left( 3\right) }
\chi_{\textrm{CNOT}}^{(12)}R_{y}^{\left( 1\right) } $ \\
9 & $R_{x}^{\left( 2\right) } \chi_{\textrm{CNOT}}^{(12)}R_{y}^{\left( 1\right) } $ \\
& \\
\hline
\end{tabular}%
\end{table}

In the four-qubit case there are $34$ sets of MUBs with different
factorizations, which are labeled by
$(n_{f},n_{b},n_{t},n_{bb},n_{nf})$, with
$n_{f}+n_{b}+n_{t}+n_{bb}+n_{nf}=2^4+1$. Then, there are $n_{f}$
factorized basis; $n_{b}$ basis with bipartite entanglement,
$|\psi \rangle _{ij}|\varphi \rangle _{k}|\varphi \rangle _{l}$;
$n_{t}$ basis with tripartite entanglement, $|\psi \rangle
_{ijk}|\varphi \rangle _{l}$; $n_{bb}$ basis with two bipartite
subsystems, $|\psi \rangle _{ij}|\varphi \rangle _{kl}$; and
$n_{nf}$ non factorized basis, where $i \neq j \neq k \neq l$ and
$i,j,k,l=1,2,3,4$. All such sets can be obtained in a regular way
by applying finite phase-space methods \cite{Klimov07}. We remark
that always there exists the so called \textquotedblleft
standard\textquotedblright\ set of MUBs, which is related to rays
in the finite phase space and can be easily constructed starting
with two classes of operators containing either $\hat{\sigma}_{z}$
or $\hat{\sigma}_{x}$ operators \cite{Klimov2005b}. The 4 qubit
case is essentially different from the above discussed $3$ qubit
case. The main difference consists in that now there exist two
locally nonequivalent completely non-factorized states. Such
states are isomorphic to the so-called graph states
\cite{Raussendorf2001}. The first type, A, are isomorphic to the
eigenstates of the set $\{\hat{\sigma}_{z}
\hat{\sigma}_{x}\hat{I}\hat{I}$,
$\hat{\sigma}_{x}\hat{\sigma}_{z}\hat{\sigma}_{x}\hat{I}$,
$\hat{I}\hat{\sigma}_{x}\hat{\sigma}_{z}\hat{\sigma}_{x}$,
$\hat{I}\hat{I}\hat{\sigma}_{x}\hat{\sigma}_{z}\}$. The second
type,B, are isomorphic to the eigenstates of the set
$\{\hat{\sigma}_{z}\hat{\sigma}_{x}\hat{I}\hat{I}$,
$\hat{\sigma}_{x}\hat{\sigma}_{z}\hat{\sigma}_{x}\hat{\sigma}_{x}$,
$\hat{I}\hat{\sigma}_{x}\hat{\sigma}_{z}\hat{I}$,
$\hat{I}\hat{\sigma}_{x}\hat{I}\hat{\sigma}_{z}\}$. Nevertheless,
both types of sets in the optimal decomposition are obtained by
applying $3$ CNOT gates only. Taking into account that MUBs with
factorizations $(1,3)$ and $(2,2)$ can be generated with $2$ CNOT
gates, we realize that the optimum set of MUBs corresponds to the
factorization structure $(0,0,12,2,3)$, which only contains type A
graph states. The complexity of such a set is $C\propto 37$ while,
for instance, the standard set of MUBs, $(3,0,0,2,12)$, has a
complexity $C\propto 40$.

The optimum set of MUBs corresponds to a set of $\allowbreak 255$ disjoint
operators which are arranged in a table consisting of 17 lines, so that each
line contains 15 commuting operators \cite{Klimov2005a}. The whole table can
be obtained from only 8 elements arranged in two lines of commuting
operators,
\begin{equation}
\begin{tabular}{|l|l|l|l|}
\hline
$\hat{1}\hat{\sigma}_{z}\hat{\sigma}_{x}\hat{\sigma}_{y}$ & $\hat{\sigma}%
_{x}\hat{\sigma}_{x}\hat{1}\hat{\sigma}_{x}$ & $\hat{1}\hat{\sigma}_{z}\hat{%
\sigma}_{y}\hat{\sigma}_{z}$ & $\hat{\sigma}_{y}\hat{\sigma}_{z}\hat{\sigma}%
_{z}\hat{\sigma}_{x}$ \\ \hline
$\hat{1}\hat{1}\hat{\sigma}_{z}\hat{1}$ & $\hat{\sigma}_{y}\hat{\sigma}%
_{x}\hat{1}\hat{\sigma}_{x}$ & $\hat{\sigma}_{z}\hat{\sigma}_{z}\hat{\sigma}%
_{z}\hat{1}$ & $\hat{\sigma}_{x}\hat{\sigma}_{x}\hat{1}\hat{\sigma}_{z}$ \\
\hline
\end{tabular}%
\end{equation}%
with factorization $(4)$ and $(1,3)$ for the first and second row,
respectively. All other operators of the above table can be generated by
using the following rule: $A_{r,c}=A_{r,c-3}\ast A_{r,c-4}$ and
$A_{r,c}=A_{2,c}\ast A_{1,r+c-3}$ for $r>2$, where the index $r$ ($c$) labels the $r$-th
MUB ($c$-th operator which has the basis states as eigenvectors of the $r$-th MUB), and
$r$ ($c$) goes from $1$ to $2^{4}+1$ ($1$ to $2^{4}-1 $). Here, the sums are modulo $15$.
The four qubit case is somewhat special, because the standard
table contains two bi-factorized bases $(2,2)$ and so there is no substantial difference
with respect to the optimum set. This apparently does not occur for larger number of
qubits, i.e, the standard table for $N>4$ would contain $2^{N}-2$ completely
non-factorized bases. The decompositions for the optimal set of MUBs is given in Table
\ref{opt4}.
\begin{table}[tbp]
\caption{The decompositions on single logic gates and controlled NOT gates for the
optimal set of MUBs $(0,0,12,2,3)$ in case of $4$-qubits.} \label{opt4}
\begin{tabular}{|c|c|}
\hline\hline
Basis & Gate operations \\ \hline\hline
\multicolumn{1}{|l|}{1} & $R_{x}^{\left( 4\right)
}\chi_{\textrm{CNOT}}^{(14)}\chi_{\textrm{CNOT}}^{(12)}R_{x}^{\left( 1\right) }$ \\
\multicolumn{1}{|l|}{2} & $R_{x}^{\left( 2\right)
}\chi_{\textrm{CNOT}}^{(31)}R_{x}^{\left(
3\right) }\chi_{\textrm{CNOT}}^{(41)}R_{x}^{\left( 4\right) }$ \\
\multicolumn{1}{|l|}{3} & $R_{y}^{\left( 2\right) }R_{x}^{\left( 1\right)
}\chi_{\textrm{CNOT}}^{(13)}\chi_{\textrm{CNOT}}^{(14)}R_{y}^{\left( 1\right) }$ \\
\multicolumn{1}{|l|}{4} & $R_{x}^{\left( 4\right) }R_{y}^{\left( 2\right)
}\chi_{\textrm{CNOT}}^{(12)}R_{x}^{\left( 1\right) }\chi_{\textrm{CNOT}}^{(13)}R_{y}^{\left( 1\right) }$ \\
\multicolumn{1}{|l|}{5} & $R_{y}^{\left( 1\right) }R_{x}^{\left( 2\right)
}\chi_{\textrm{CNOT}}^{(34)}R_{x}^{\left( 3\right) }\chi_{\textrm{CNOT}}^{(24)}R_{x}^{\left( 2\right) }$ \\
\multicolumn{1}{|l|}{6} & $R_{y}^{\left( 4\right) }R_{x}^{\left( 3\right)
}\chi_{\textrm{CNOT}}^{(23)}R_{x}^{\left( 2\right) }\chi_{\textrm{CNOT}}^{(31)}R_{x}^{\left( 3\right) }$ \\
\multicolumn{1}{|l|}{7} & $R_{y}^{\left( 1\right)
}\chi_{\textrm{CNOT}}^{(13)}\chi_{\textrm{CNOT}}^{(12)}R_{x}^{\left( 1\right) }$ \\
\multicolumn{1}{|l|}{8} & $R_{y}^{\left( 3\right) }R_{y}^{\left( 1\right)
}\chi_{\textrm{CNOT}}^{(14)}R_{x}^{\left( 1\right) }\chi_{\textrm{CNOT}}^{(12)}R_{y}^{\left( 1\right) }$ \\
\multicolumn{1}{|l|}{9} & $R_{x}^{\left( 1\right) }R_{x}^{\left( 2\right)
}R_{x}^{\left( 4\right) }\chi_{\textrm{CNOT}}^{(23)}\chi_{\textrm{CNOT}}^{(24)}R_{y}^{\left( 2\right) }$ \\
\multicolumn{1}{|l|}{10} & $R_{x}^{\left( 1\right)
}\chi_{\textrm{CNOT}}^{(34)}R_{y}^{\left(
3\right) }\chi_{\textrm{CNOT}}^{(14)}R_{x}^{\left( 1\right) }$ \\
\multicolumn{1}{|l|}{11} & $R_{x}^{\left( 3\right)
}\chi_{\textrm{CNOT}}^{(24)}R_{y}^{\left(
2\right) }\chi_{\textrm{CNOT}}^{(14)}R_{x}^{\left( 1\right) }$ \\
\multicolumn{1}{|l|}{12} & $R_{y}^{\left( 2\right) }R_{y}^{\left( 3\right)
}\chi_{\textrm{CNOT}}^{(23)}\chi_{\textrm{CNOT}}^{(24)}R_{y}^{\left( 2\right) }$ \\
\multicolumn{1}{|l|}{13} & $\chi_{\textrm{CNOT}}^{(13)}R_{y}^{\left( 1\right)
}\chi_{\textrm{CNOT}}^{(24)}R_{x}^{\left( 2\right) }$ \\
\multicolumn{1}{|l|}{14} & $R_{y}^{\left( 2\right)
}\chi_{\textrm{CNOT}}^{(14)}R_{y}^{\left(
1\right) }\chi_{\textrm{CNOT}}^{(23)}R_{y}^{\left( 2\right) }$ \\
\multicolumn{1}{|l|}{15} & $R_{y}^{\left( 4\right) }R_{x}^{\left( 1\right)
}\chi_{\textrm{CNOT}}^{(34)}R_{x}^{\left( 3\right)
}\chi_{\textrm{CNOT}}^{(32)}\chi_{\textrm{CNOT}}^{(31)}R_{x}^{\left(
3\right) }$ \\
\multicolumn{1}{|l|}{16} & $R_{x}^{\left( 3\right)
}\chi_{\textrm{CNOT}}^{(13)}R_{x}^{\left( 1\right)
}\chi_{\textrm{CNOT}}^{(34)}R_{x}^{\left( 3\right)
}\chi_{\textrm{CNOT}}^{(12)}R_{y}^{\left( 1\right)
}$ \\
\multicolumn{1}{|l|}{17} & $R_{x}^{\left( 2\right) }R_{x}^{\left( 3\right)
}\chi_{\textrm{CNOT}}^{(12)}R_{y}^{\left( 1\right)
}\chi_{\textrm{CNOT}}^{(23)}\chi_{\textrm{CNOT}}^{(24)}R_{x}^{\left( 2\right) }$
\\ \hline
\end{tabular}
\end{table}

The situation is quite different in the 5 qubit case, in which
there exist at least 9000 non-isomorphic sets of MUBs with
different factorizations, which are labelled by
$(n_{f},n_{b},n_{t},n_{bb},n_{f},n_{bt},n_{nf})$, where $n_{f}$
denotes the number of bases with four-particle entanglement,
$|\psi \rangle _{ijkl}|\varphi \rangle _{n}$, and $n_{bt}$ the
number of bases with bipartite and tripartite entanglement, $|\psi
\rangle _{ijk}|\varphi \rangle _{lm}$, with $i \neq j \neq k \neq
l \neq m$ and $i,j,k,l,m=1,2,3,4,5$. In case of 5 qubits there are
four locally nonequivalent completely non-factorized states
\cite{Hein2004}. In this case the standard table does not contain
partially factorized bases and has the structure $\left(
3,0,0,0,0,0,30\right) $, and it is given by:
\begin{equation*}
\begin{tabular}{|l|l|l|l|l|}
\hline $\hat{1}\hat{1}\hat{\sigma}_{z}\hat{\sigma}_{z}\hat{1}$ &
$\hat{\sigma}_{z}\hat{\sigma}_{z}\hat{\sigma}_{z}\hat{1}\hat{\sigma}_{z}$ &
$\hat{1}\hat{1 }\hat{1}\hat{1}\hat{\sigma}_{z}$ &
$\hat{1}\hat{\sigma}_{z}\hat{1}\hat{\sigma}_{z}\hat{1}$ &
$\hat{1}\hat{1}\hat{1}\hat{\sigma}_{z}\hat{1}$ \\ \hline
$\hat{1}\hat{1}\hat{\sigma}_{x}\hat{\sigma}_{x}\hat{1}$ & $\hat{\sigma}
_{x}\hat{\sigma}_{x}\hat{\sigma}_{x}\hat{1}\hat{\sigma}_{x}$ &
$\hat{1}\hat{1}\hat{1}\hat{1}\hat{\sigma}_{x}$ &
$\hat{1}\hat{\sigma}_{x}\hat{1}\hat{\sigma}_{x}\hat{1}$ &
$\hat{1}\hat{1}\hat{1}\hat{\sigma}_{x}\hat{1}$ \\ \hline
\end{tabular}.
\end{equation*}
The set of MUBs corresponds to $\allowbreak 1023$ disjoint operators which are arranged
in a table consisting of 33 lines, so that each line contains 31 commuting operators. All
the other projectors of this table are obtained
by using the following rule: $A_{r,c+18}=A_{r,c}\ast A_{r,c+1}$ and $%
A_{r,c}=A_{1,c}\ast A_{2,r+c-2}$, for $r>2$, where the sums are modulo $31$. This table
contains $30$ non-factorized bases among which there are three types corresponding to
different graphs: $6$ of type B, $18$ of type C and $6$ of type D (graphs 6, 7 and 8 in
Fig.4 \cite{Hein2004}). It results that 4 CNOT gates are required to generate bases of
type B and C, nevertheless the minimum number of CNOT gates to obtain the type D graphs
is 5. This means that one needs 126 non-local operations to generate the whole set of
MUBs, with the corresponding complexity $C\propto 126$. The optimum set has the structure
$\left( 0,0,1,3,10,2,17\right) $ corresponding to a complexity $C\propto 112$, and
contains one non-factorized basis of type B and $16$ bases of type C. In this case the
set of operators needed for generating the whole table is
\begin{small}
\begin{equation*}
\begin{tabular}{|l|l|l|l|l|}
\hline
$\hat{\sigma}_{x}\hat{\sigma}_{z}\hat{\sigma}_{z}\hat{\sigma}_{x}\hat{%
\sigma}_{z}$ & $\hat{\sigma}_{x}\hat{\sigma}_{z}\hat{1}\hat{\sigma}_{y}\hat{%
\sigma}_{y}$ & $\hat{\sigma}_{z}\hat{\sigma}_{z}\hat{1}\hat{1}\hat{\sigma}%
_{x}$ & $\hat{\sigma}_{y}\hat{\sigma}_{z}\hat{\sigma}_{z}\hat{\sigma}_{y}%
\hat{\sigma}_{z}$ & $\hat{\sigma}_{x}\hat{1}\hat{\sigma}_{z}\hat{\sigma}_{x}%
\hat{\sigma}_{z}$ \\ \hline
$\hat{1}\hat{\sigma}_{y}\hat{\sigma}_{y}\hat{\sigma}_{x}\hat{\sigma}_{z}$
& $\hat{\sigma}_{z}\hat{\sigma}_{y}\hat{\sigma}_{y}\hat{\sigma}_{z}\hat{%
\sigma}_{x}$ & $\hat{\sigma}_{z}\hat{\sigma}_{z}\hat{1}\hat{\sigma}_{z}\hat{%
\sigma}_{z}$ & $\hat{\sigma}_{z}\hat{1}\hat{\sigma}_{x}\hat{\sigma}_{x}\hat{%
\sigma}_{x}$ & $\hat{1}\hat{\sigma}_{z}\hat{\sigma}_{x}\hat{1}\hat{1}$ \\
\hline
\end{tabular}.
\end{equation*}%
\end{small}
The rule for generating the other projectors is the same as in the
standard table. For this table the decompositions on single logic
gates and CNOT gates can be obtained in the same way as for the
above discussed cases.

We have studied the problem of optimal tomographic reconstruction
of a density operator of systems of $N$ cold trapped ions. The
optimality of a given set of MUBs is essentially defined in terms
of the minimum number of required conditional operations. We have
given an explicit form of operations to generate the optimal set
of MUBs for the case of three and four qubits, and the
factorization operations for the five qubit case can be obtained
in the same way. In the case of larger numbers of qubits, a
generic procedure allowing generation of the whole set of MUBs is
also available. It basically consists in finding all the possible
strings of $2^{N}-1$ commuting operators (using explicit
geometrical construction) with a subsequent separation onto the
$2^{N}+1$ disjoint sets \cite{Bandyopadhyay}. Eigenstates of such
commuting operators in each disjoint set generate one of the
corresponding MUBs \cite{Klimov07}. This reconstruction scheme is
valid for any physical setup, where non-local operations between
neighboring qubits have the same fidelity as between distant
qubits. This is satisfied in case of trapped ions because a CNOT
gate, between any pair of ions, is implemented by using the center
of mass motion as a data bus. If the physical implementation does
not meet this requirement, the optimal MUBs to be used for
reconstructing the state must have to be determined by considering
the fidelities between non-neighboring qubits.

\textbf{Acknowledgment} We thanks to J. Eschner and M. Yang for
useful comments. This work was supported by Grants Milenio ICM
P06-67F, FONDECyT 1061046, and CONACyT 45704. A.F. thanks to
CONICyT for scholarship support.


\begin{thebibliography}{99}
\bibitem{Bennett2000} C.H. Bennett and D.P. DiVincenzo, Nature \textbf{404}, 247
(2000).

\bibitem{Fano1957} U. Fano, Rev. Mod. Phys. \textbf{29}, 74 (1957).

\bibitem{Blatt2005} H. H\"{a}ffner \textit{et al.}, Nature \textbf{438}, 643 (2005).

\bibitem{Cirac1995} J.I. Cirac and P. Zoller, \prl \textbf{74}, 4091 (1995).

\bibitem{Blatt2004} C. F. Roos \textit{et al.}, \prl \textbf{92}, 220402
(2004); M. Riebe \textit{et al.}, \prl \textbf{97}, 220407
(2006); M. Riebe \textit{et al.}, New J. Phys. \textbf{9}, 211
(2007).

\bibitem{Mitchell2003} M.W. Mitchell \textit{et al.}, \prl \textbf{91},
120402 (2003).

\bibitem{Bouland2003} N. Boulant \textit{et al.}, \pra \textbf{67}, 042322
(2003).

\bibitem{Hradil1997} Z. Hradil, \pra \textbf{55}, R1561 (1997).

\bibitem{James2001} D.F.V. James \textit{et al.}, \pra \textbf{64}, 052312
(2001).

\bibitem{Wootters1989} W.K. Wooters, B.D. Fields, Ann. Phys. \textbf{191},
363 (1989).

\bibitem{Englert2004} J. \v{R}eh\'{a}\v{c}ek, B.G. Englert, and D. Kaszlikowski, Phys. Rev. A \textbf{70}%
, 052321 (2004).

\bibitem{Ling2006} A. Ling, K.P. Soh, A. Lamas-Linares, and C. Kurtsiefer, Phys. Rev. A
\textbf{74}, 022309 ( 2006).

\bibitem{Monroe1995} C. Monroe \textit{et al.}, \prl\textbf{75}, 4714 (1995).

\bibitem{Lawrence2002} J. Lawrence, \v{C}. Brukner, and A. Zeilinger, \pra%
\textbf{65}, 032320 (2002).

\bibitem{Klimov2005a} J. L. Romero \textit{et al.}, \pra \textbf{72}, 062310
(2005).

\bibitem{Bandyopadhyay} S. Bandyopadhyay \textit{et al.}, Algorithmica
\textbf{38}, 512 (2002).

\bibitem{Raussendorf2001} R. Raussendorf and H. J. Briegel, \prl\textbf{86},
5188 (2001).

\bibitem{Hein2004} M. Hein, J. Eisert, and H. J. Briegel, \pra \textbf{69},
062311 (2004).

\bibitem{transfor} These rotations are given in Ref. \cite{Blatt2004} and
their action logical states is given by: $R_{x}^{(j)} \left\vert 0_{j}\right\rangle
=\frac{1}{\sqrt{2}}\left( \left\vert 0_{j}\right\rangle -i\left\vert 1_{j}\right\rangle
\right)$, $R_{x}^{(j)} \left\vert 1_{j}\right\rangle
=\frac{1}{\sqrt{2}}\left(\left\vert1_{j}\right\rangle -i\left\vert 0_{j}\right\rangle
\right) $; $R_{y}^{(j)} \left\vert 0_{j}\right\rangle =\frac{1}{\sqrt{2}}\left(
\left\vert 0_{j}\right\rangle +\left\vert 1_{j}\right\rangle \right)$, $R_{y}^{(j)}
\left\vert 1_{j}\right\rangle =\frac{1}{\sqrt{2}}\left(\left\vert
1_{j}\right\rangle-\left\vert 0_{j}\right\rangle \right)$; and $R_{z}^{(j)} \left\vert
0_{j}\right\rangle = i\left\vert 0_{j}\right\rangle$, $R_{z}^{(j)}\left\vert
1_{j}\right\rangle =\left\vert 1_{j}\right\rangle$.

\bibitem{Childs2000} A.M. Childs, I.L. Chuang, \pra \textbf{63}, 012306
(2000).

\bibitem{Klimov2005b} A.B. Klimov, L.L. Sánchez-Soto, H. de Guise, J.
Phys. A \textbf{38}, 2747 (2005); A.B. Klimov \textit{et al.}, J. Phys. A \textbf{39},
14471 (2006).

\bibitem{Werner2002} D. Schlingemann and R.F. Werner, \pra \textbf{65},
012308 (2002).

\bibitem{Guang2006} J.-M. Cai, Z.-W. Zhou, and G.-C. Guo,
quant-ph/0609186 (2006).

\bibitem{Klimov07} G. Bjork \textit{et al}. JOSA B, 24, 371 (2007); A.B.
Klimov \textit{et al.} LNCS \textbf{4547}, 333 (2007).
\end{thebibliography}
\end{document}